\documentstyle[12pt,epsf]{article}

\def\frc{f}

\def\comment#1{}

\begin{document}
\date{}
\title {\textsf{Flashing Dark Matter\\---\\
Gamma-Ray Bursts from  Relativistic Detonations \\of
Electro-Dilaton Stars}}
\author { V.~Folomeev,
V.~Gurovich\thanks{email: astra@freenet.kg \
  Physics Institute of NAN KR,
  265 a, Chui str., Bishkek, 720071,  Kyrgyz Republic},
\,H. Kleinert\thanks{email: kleinert@physik.fu-berlin.de \
Inst. f. Theor. Physik, Freie Univ. Berlin}
 \  and
H.-J. Schmidt\thanks{email: hjschmi@rz.uni-potsdam.de \
 Inst. f. Math. Univ. Potsdam, Am Neuen Palais 10, D-14469 Potsdam,
Germany}}

\maketitle

\begin{abstract}
We speculate that the universe  is filled with stars
composed of electromagnetic and dilaton fields
which are the sources of the powerful gamma-ray bursts
impinging upon us from all directions
of the universe.  We calculate    soliton-like solutions
of these fields  and  show that their  energy   can be converted
into a  relativistic plasma in  an explosive way.
As in classical detonation theory  the conversion proceeds by
 a  relativistic  self-similar solution for a spherical detonation wave
which extracts the energy from the scalar field
via a plasma in the wave front in the atmosphere of the star.
\end{abstract}

\section{Introduction}

Many modern  theories of the  universe
 assume the existence of various types of   scalar fields. Such fields could
explain the  recently discovered acceleration  of the expansion of the universe
(see e.g.  Ref.~\cite{ref:Sahni}), or the   formation of clustered systems leading to
gravitational walls for galaxies and galaxy clusters. If such fields really exist,
 the universe could  contain  many   compact star-like  configurations of large
 total mass, called scalar stars.

Among the many possible scalar fields,
the dilaton field   has a special theoretical appeal.
It  couples in a unique minimal way to electromagnetism
to make the Maxwell action dimensionless.
This coupling leads to star-like objects
which are composed of scalar and
electromagnetic fields.  That such objects can exist
was pointed out by many authors~\cite{ref:Ruf,ref:Pir,ref:Miel}.
We argue that such electro-dilaton stars
may be responsible  for the strong gamma-ray bursts observed in the universe.

At present there exists    no  simple
conventional explanation for  the origin of these
events reaching us isotropically  from all directions of the universe.
For an comprehensive discussion of the subject,
in particular  for the failures of most theories,
see the article by Ruffini \cite{RU}.
 At the same   time,  there exist  models for the early
universe  where   an  initial state  of  large volume of relativistic
plasma quickly expands as  a spherical
wave, like a critical bubble  in an overheated liquid,
causing a  decay of a false vacuum, and
creating the universe from such a bubble~\cite{ref:Col}.
The electro-dilaton stars  can supply such an explanation.
We   assume that   the  dark  matter in the universe
contains a multitude  of such objects, whose total
 mass  exceeds by far the total mass of  luminous
matter and is responsible  for  the large-scale structure of the
universe. The luminous  matter concentrates in the gravitational
potential wells of the scalar fields.

Let us imagine that collision of relativistic  particles produce
a fireball of critical size. Such bubbles have been investigated as
possible  triggers of phase transitions
in the early universe \cite{ref:Col}.  In the context of
gamma-ray bursts, similar assumptions have been made in
Ref.~\cite{ref:Post}.
The critical bubble forms the seed for
 transferring   effectively scalar  fields  into
 pairs of elementary particles and their antiparticles.
The transfer may be initiated  by fast oscillation of a
field on the outer boundary of the  fireball.  The process causes
 a relativistic detonation.  In a conventional detonation,    chemical
energy is converted into kinetic energy.
In a relativistic detonation of an electro-dilaton star,  it is the
energy of
the electric and scalar fields  which is rapidly converted  in
particle-antiparticle pairs.
The resulting fireball expands with relativistic  velocity and
will therefore not depend on the weak gravitational fields of a
Newtonian configuration. We may thus study the process within special
relativity.

\section{Relativistic Detonation}
\label{sec1}
We begin by deducing the self-similar solutions for a spherical
relativistic detonation which goes over to  the well-known
 Zeldovich  solution
in the limit  of small velocities~\cite{ref:Land}.
The set of equations of relativistic hydrodynamics is conveniently
 described  in a spherical coordinate system $r,\Theta,\phi$. If $v$
denotes the     radial velocity of the plasma
 in  three-dimensions~\cite{ref:Land,ref:Baum} and
$\varepsilon$ is energy density, $p$ - pressure,
the equation of motion reads
\begin{equation}
\label{eq1}
\frac{1}{\gamma ^{2}}\left( \frac{\partial v}{\partial \tau }+v\frac{%
\partial v}{\partial r}\right)
+\frac{1}{W}\left( \frac{\partial p}{\partial
 r}+v\frac{\partial p}{\partial \tau }\right) =0
\end{equation}
while energy conservation requires that
\begin{equation}
\label{eq2}
\frac{1}{W}\left[ \frac{\partial \varepsilon }{\partial \tau }+v\frac{%
\partial \varepsilon }{\partial r}\right]
+\frac{1}{\gamma ^{2}}\left( \frac{%
\partial v}{\partial r}+v\frac{\partial v}{\partial \tau }\right)
+\frac{2v}{%
r}=0,
\end{equation}
where $\gamma ^{2}\equiv 1-v^{2}$, $W=\varepsilon +p$,\ and $c=1$.
As  in the  nonrelativistic case, the motion of the plasma behind
the   detonation front  is considered as isentropic,
such that
(\ref{eq1}) and (\ref{eq2}) are the only relevant equations.

The pairs of relativistic particles and antiparticles  created
 behind the wave front   generate a high-temperature plasma
with the equation of state:
\begin{equation}
\label{eq3}
p=c_s  ^{2}\,\varepsilon ;\,\,\,\,\,\,\,\,\,c_s  ^{2}=\left( \frac{%
\partial p}{\partial \varepsilon }\right) _{S}=\frac{1}{3},
\end{equation}
where $c_s  $\ is the sound velocity. As  in
an ordinary spherical detonation problem,
we  search for a solution  depending on the self-similar variable
\begin{equation}
\label{eq4}
 \xi =r/\tau ,
\end{equation}
in which the differential equations  (\ref{eq1}) and (\ref{eq2}) reduce to
 ordinary differential equations,   which can be combined to
the equation for  $v$:
\begin{equation}
\label{eq5}
\frac{dv}{d\xi }\left[ \frac{1}{c_s  ^{2}}\left( \frac{v-\xi }{1-v\xi }%
\right) ^{2}-1\right] =\frac{2v}{\xi }\frac{\gamma ^{2}}{1-v\xi }.
\end{equation}
In the nonrelativistic limit  where $v(\xi ),\,\xi \ll 1$, this
equation reduces to a textbook
equation~\cite{ref:Land}. The qualitative
analysis of the relativistic equation
(\ref{eq5}) is
similar to the one in the textbook: The solutions for $v$\ and $%
\varepsilon $\ have a diverging   derivative at the wave front.
Such singularity of a derivative is defined by the following:
the expression in parentheses is  velocity of  current of plasma
in relation to the wave front. According to the theory of detonation,
this velocity is equal to sound velocity $c_s$. Therefore,
at approaching to the wave front the expression in
brackets tends to zero. The coordinate of the wave front $\xi=v_d$
(where $v_d$ is a velocity of detonation wave in central frame)
 and the velocity of gas behind the wave in  laboratory
frame $v_g$ is uniquely determined from conservation
laws at the wave front.

\bigskip
\bigskip
\bigskip
\bigskip
\begin{figure}[tbhp]
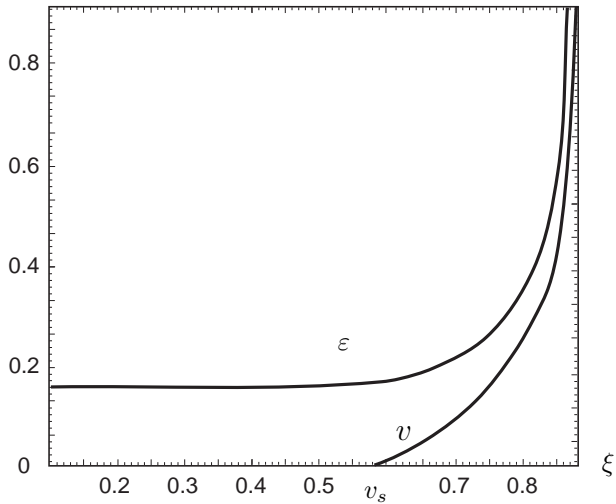

~\vspace{1cm}
~\\
\input  fig1.tps
\caption[]{Dependence of the energy density $\varepsilon$
and velocity $v$ of plasma on the self-similar variable $\xi$
behind the front of detonation wave.
}
\label{@Fig1}\end{figure}
The discussion of possible mechanisms of  ``recycling'' of the
field behind the wave front into relativistic plasma is considered
below. Let us specify here the following estimates of the  value of
${v_d}$\ and energy density behind the detonation wave.
Consider a scalar field in the simplest
form, with an energy-momentum tensor
\begin{equation}
\label{eq6}
T_{i}^{k}=\varphi _{,i}\varphi ^{,k}-\delta _{i}^{k}\left[ \frac{1}{2}%
\varphi _{,\mu }\varphi ^{,\mu }-V(\varphi )\right]
,\,\,\,\,\,\,\,\,\,V(\varphi )=m^{2}\varphi ^{2}/2.
\end{equation}
In the ``scalaron" regime where it undergoes
 fast oscillations of frequency $m$, i.e. where
$\varphi (r,t)=a(r)\sin mt$, the spatial gradients
  of the field can be ignored and
 the energy density in the laboratory frame) is
\begin{equation}
\label{eq7}
\varepsilon _{f}\approx m^{2}a^{2}/2.
\end{equation}
The expression for ${v_d}$\ and the energy density of the plasma
behind the wave front are
determined from the conservation
laws $T_{0}^{1}({\rm field})=T_{0}^{1}({\rm plasma})$ and $
T_{1}^{1}({\rm field})=T_{1}^{1}({\rm plasma})$
 for an observer moving with     the wave front. Recall
  that the plasma is emitted from    the
wave front with the velocity $c_s  $. Hence
\begin{equation}
\label{eq8}
v_g=c_s; \quad {{v_d}}=\frac{2c_s  }{1+c_s  ^{2}};\,\,\,\,\,\varepsilon _{p}=\frac{2}{%
1-c_s  ^{2}}\varepsilon _{f}.
\end{equation}
Since  a relativistic plasma  has $c_s  =1/\sqrt{3}$, we find
${{v_d}}=\sqrt{3}%
/2$ and $\varepsilon _{p}=3\,\,\varepsilon _{f}\,$. The
self-similar solutions for this case are presented in Fig.~\ref{@Fig1}.

We now turn to the   mechanism of transition of the field energy into
a relativistic plasma.

\section{Electro-Dilaton Wave}

So far, the description of the  conversion  of
the energy of a  scalar field
into relativistic plasma at the front of  ``detonation" wave
 is purely phenomenological.  The physical  properties
 of the front of the detonation wave are
completely determined by the conservation laws for a general
energy-momentum tensor $T_{i}^{k}$.

One  specific  mechanism of such a conversion
was considered in~\cite{ref:Tkach}  based on an
analogy  with a laser.  Here we shall consider an alternative
 possibility where the relativistic plasma and radiation
are obtained from the energy of a  dilaton field.

The star-like configurations for such field with strong and weak
(Newtonian) gravitational field have been
considered before~\cite{ref:Tao,ref:Greg,ref:Mat,ref:Fiz}. As
 in Section \ref{sec1},  we shall discuss only
the case of   a weak gravitational field and   treat
the  problem within special relativity.

The Lagrangian density of a system of dilaton and electromagnetic fields
is~\cite{ref:Greg,ref:Mat}:
\begin{equation}
\label{lagr}
L=2\Phi_{,i} \Phi^{,i}-\zeta (F_{lm}F^{lm})e^{-2\alpha\Phi}.
\end{equation}
The parameter $\zeta$ can have the values $\pm 1$\footnote{our choice
is $\zeta = -1$}
as will be explained  below. The normalization of the fields
is the same as in \cite{ref:Greg,ref:Mat}.
 The  unique
interaction of the scalar field $\Phi$ with an electromagnetic field
required by scale   invariance    allows for a
nontrivial combined electro-dilaton
configuration. In the front of the   detonation wave,
the electromagnetic field   reduces
the dilaton field strength by   dissipation.
 Depending on the intensity of the electric field,
the dissipation may be due
to the creation of pairs of particles and antiparticles
and to a heating  of the  plasma. The combined process---generation of an
electromagnetic field and its subsequent dissipation---supplies
the energy to the   front of the detonation wave. Let us study the process
in a simple plane-wave configuration.   The equations
for $\Phi$ and $F_{ik}$ following from Lagrangian~(\ref{lagr})
are
\begin{eqnarray}
\label{cons1}
\left[e^{-2\alpha\Phi}F^{ik}\right]_{;k}&=&0,
\\
\quad \Phi_{;i}^{;i}&=&-\frac{\alpha
\zeta}{2}e^{-2\alpha\Phi}(F_{lm}F^{lm}).
\label{cons1b}
\end{eqnarray}
The system is supplemented by the missing Maxwell equations
(electromagnetic versions of the Bianchi identities):
\begin{equation}
\label{maxv}
e^{iklm}F_{kl,m}=0.
\end{equation}
The total energy-momentum tensor  associated with the  Lagrange density
(\ref{lagr}) is
\begin{equation}
\label{ener}
T_{i}^{k}=2\Phi_{,i}\Phi^{,k}-2 \zeta e^{-2\alpha\Phi}
F_{il}F^{lk}-\frac{1}{2}\,\delta_{i}^{k}[2(\Phi_{,l}\Phi^{,l})
-\zeta e^{-2\alpha\Phi}(F_{lm}F^{lm})].
\end{equation}

\section{Longitudinal Electro-Dilaton Wave}

We now show that there exist plane electro-dilaton waves
 travelling along the $x$-axis in which the
electric field has a component $F^{10}=E_x$
with all other components vanishing.
 Whereas Eq. (\ref{maxv}) is fulfilled
trivially,  Eq.~(\ref{cons1}) yields
\begin{eqnarray}
\label{cons2}
(e^{-2\alpha\Phi}E_x)_{,\tau}&=&0; \quad \quad \quad
(e^{-2\alpha\Phi}E_x)_{,x}=0.
\end{eqnarray}
Thus we find a constant of motion
\begin{eqnarray}
\label{cons2a}
 e^{-2\alpha\Phi}E_x=E_0={\rm const.}
\end{eqnarray}
The  quadratic field combination
\begin{eqnarray}
\label{cons2b}
I=F_{lm}F^{lm}=-2 E_{x}^{2}
\end{eqnarray}
has the same  negative value
 in all systems of special relativity. For this reason we
follow Ref.~\cite{ref:Greg}
in choosing the parameter
$\zeta=-1$ in  (\ref{lagr}).

The other field equation (\ref{cons1b}) becomes
\begin{equation}
\label{wave}
\frac{\partial\Phi}{\partial\tau^2}-\frac{\partial\Phi}{\partial x^2}=\alpha
E_{0}^{2}e^{2\alpha\Phi}.
\end{equation}
or, introducing new
variables $2\alpha E_0 x \rightarrow x$, $2\alpha E_0 \tau \rightarrow \tau$
and function $\psi=2 \alpha \Phi$, we have
\begin{equation}
\label{wave1}
\frac{\partial\psi}{\partial\tau^2}-\frac{\partial\psi}{\partial x^2}=e^{\psi}/2.
\end{equation}
This is one of form of Liouville equation~\cite{ref:Lio}. Its
 possesses a steady-state wave solution
which, after the redefinition of the variables
\begin{equation}
\label{redef}
\xi=x-u \tau,~
\quad \eta=\xi/\sqrt{1-u^2},
\end{equation}
takes the form
\begin{equation}
\label{eqn8}
\frac{d^2\psi}{d\eta^2}=-\frac{1}{2}e^{\psi}.
\end{equation}
The first integral of this equation  leads to the differential equation
\begin{equation}
\label{eqn9}
\left(\frac{d\psi}{d\eta}\right)^2=1-e^{\psi}
\end{equation}
whose  solution
\begin{equation}
\label{eqn10}
\psi=-2\ln\cosh(\eta/2),~~~ \quad E_x=E_0/\cosh^2(\eta/2)
\end{equation}
describes a  soliton.
Though the potential of dilaton field $\psi$ diverges at
$|\eta|\rightarrow\infty$,
where $\Phi\approx-\eta$, the   derivatives $\Phi_{,i}$ remain
finite, thus   ensuring   a finite energy density of the field for all $\eta$.
Note that in contrast  to a charged plane in electrostatics,
 the electric field of this  solution has a zero flux $E_x$ at infinity.
Asymptotically, no  electric   field is detectable.

There exists a general class  of solutions of the Liouville
equation (\ref{wave1}) containing two arbitrary functions $f_1(x-\tau),\,
f_2(x+\tau)$.  Setting
$\psi^{\ast }\equiv 2 \alpha \Phi +\ln  \alpha E_0^2$,
the solution of eq. (\ref{wave}) is
\begin{equation}
\label{solut}
\psi^{\ast }=\ln \left[\frac{16 f^{'}_{1}(x-\tau)
f^{'}_{2}(x+\tau)}{\cosh^2[f_{1}(x-\tau)+ f_{2}(x+\tau)]}\right]
\end{equation}
The  primes denote derivatives.
Especially simple solutions  are obtained for the linear
functions
\begin{eqnarray}
\label{func}
f_1(x-\tau )&=&  \gamma \,(x-\tau),   ~~~~
 \gamma \equiv \frac{1+u}{4\sqrt{1-u^2}}   ,\\
 \quad f_2(x+\tau)&=& \beta \,(x+\tau),~~~~ \beta \equiv
\frac{1-u}{4\sqrt{1-u^2}} .
\label{func2}
\end{eqnarray}
where $u$ is velocity of soliton in Eq.(\ref{redef}).

The following solutions are of special interest:
\begin{enumerate}
\item[a]$\!\!$)   Localized
solution for $E_x$  with a fixed
asymptotic  energy
density $\varepsilon(| x| \rightarrow \infty)=\varepsilon_0$,
where $E_x$ goes over into  the previous soliton
solution (\ref{eqn10}). In these solutions, the functions $f_1$ and $f_2$
in (\ref{solut}) and  their derivatives are regular.
Singular solutions of
Liouville equation are also known~\cite{ref:Jorj}. They
may lead to a local catastrophic growth of electric field, and
require special attention.

\item[b]$\!\!$)
 The main purpose of this section is to show
that at different values of the energy of electro-dilaton wave
before and behind the electric layer the electromagnetic
energy may increase in time.
A similar distributions of the energy may be created
by gravitational configuration.

Let us specify the arbitrary functions in Liouville solution as
\begin{eqnarray}
\label{func1}
f_1&=&\gamma (x-\tau)+F, \quad F=-\ln \tanh \mu (x-\tau), \\
f_2&=&\beta (x+\tau),
\end{eqnarray}
\begin{figure}[h]
~\\[2cm]
\begin{picture}(500,120)
\put(20,-300){\includegraphics{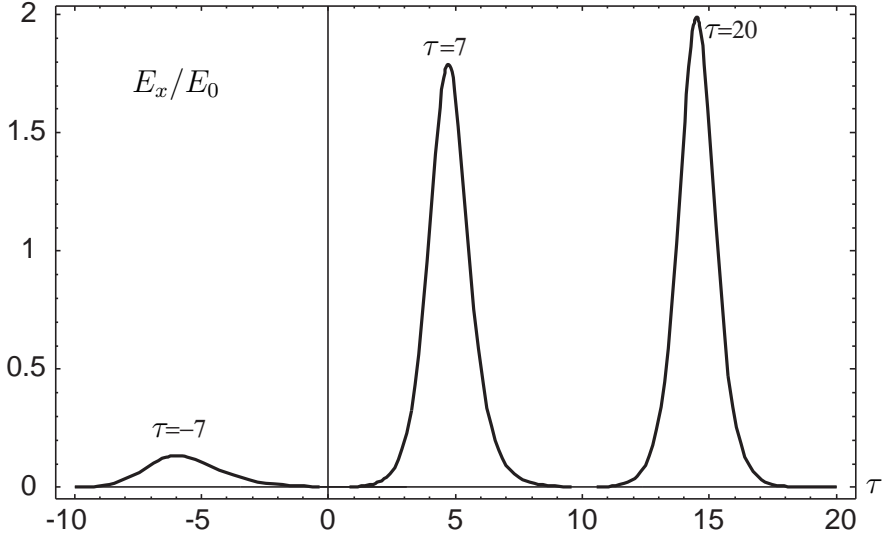}}
\put(360,-00){\normalsize $\tau $}
\put(90,21.5){\normalsize $\tau $}
\put(192.5,165.8){\normalsize $\tau $}
\put(298.9,172.5){\normalsize $\tau $}
\put(82.5,152){\normalsize $E_x/E_0 $}
\end{picture}
\caption[]{Growth of electrical soliton in wave with different
asymptotic values of energy both before and behind wave.
Solutions (\ref{elec}) with $u=1/\sqrt{3}$ and $\mu=\gamma$ are
shown for increasing times $\tau$. The  energy density in
units of $E^2_0$ is equal to $7.4$ behind and $0.2$ in front of the
wave front.
For large $\tau$ the solution gives stationary soliton with smaller value
of energy.}
\label{fig3}
\end{figure}
~\\
where $\gamma, \beta$ are determined
by  the initial velocity of the wave  from   (\ref{func})
and (\ref{func2}), and $\mu$ is an arbitrary constant.
 Now the profile of the electric field looks like
\begin{equation}
\label{elec} \frac{E_x}{E_0}=\frac{16 \beta [\gamma+\mu \tanh \mu
(\tau-x)]} {\cosh^2[(\gamma+\beta)x - (\gamma - \beta)\tau+\ln
\cosh \mu (\tau-x)]}
\end{equation}

The profile of such a wave is shown on Fig.~\ref{fig3}. Let us
assume that  the numerator of Eq. (\ref{elec}) slowly changes in
comparison with the denominator. In that case, at given $\tau$ the
maximum of electric soliton concentrates  about a zero point of
argument of $\cosh$.  As we see from
 Eq.~(\ref{func}), this region moves with velocity $u<1$,
 so that  the argument of $\tanh$ is always positive in this region. The
 numerator and thus the electric field
increase with time. This increase comes to an end
as   $\tanh[\mu(\tau-x)]$   reaches its asymptotical value 1. This is
illustrated in Fig.~\ref{fig3}. If
 instead of   $\tanh[\mu(\tau-x)]$  in (\ref{func1})
 a function  is chosen which grows without bounds,
then also the   electric field will keep growing.

The growth of the energy of the electric field will
be consumed by  dissipation.
 Its influence on the electro-dilaton wave will be considered
in Section~\ref{@DISS}.

\end{enumerate}

\section{Wave of Transverse-Magnetic Type}

Let us use the above results to study a
 wave with a transverse magnetic
wave. In this case the Maxwell equations (\ref{maxv}) are not fulfilled
 identically but yield
 a relation between transverse components $F_{ik}$. Let us select for
consideration the following nonzero components $F^{20}=E_y, F^{21}=H_z$.
Then the complete set of the equations is
\begin{eqnarray}
\label{eqn11}
\frac{\partial}{\partial\tau}(e^{-2\alpha\Phi}E_y)&+&\frac{\partial}{\partial
x}(e^{-2\alpha\Phi}H_z)=0,\\
\frac{\partial E_y}{\partial x}&+&\frac{\partial H_z}{\partial\tau}=0.
 \label{11b}
\end{eqnarray}
A travelling wave has the field components
\begin{equation}
\label{eqn12}
E_y=u H_z(\xi); \quad H_z=
H_0 e^{2\alpha\Phi}; \quad I=2 H_{0}^{2}(1-u^2)e^{4\alpha\Phi}.
\end{equation}
Since $u<1$ and $I>0$, the field is transverse magnetic.
 It dictates in (\ref{cons1})
the choice of the sign  $\zeta=1$. Further, using similar
variables as in (\ref{redef})
\begin{equation}
\label{eqn13}
\xi=x-u \tau,~ \quad \psi=2 \alpha\Phi,~ \quad \eta=2\alpha\xi| H_0|,
\end{equation}
we obtain the solutions for dilaton and electromagnetic fields
similar to (\ref{eqn10})
\begin{equation}
\label{eqn14}
\psi=-2\ln\cosh(\eta/2); \quad E_y=u H_z=u H_0/\cosh^2(\eta/2).
\end{equation}
The principal difference with respect to
(\ref{eqn10})
is the absence
of a relativistic factor $\sqrt{1-u^2}$ in the argument $ \eta $.

The   transverse magnetic wave is focused in a band of width
 $\triangle x\sim  1/\alpha| H_0|$.
Outside of this, the energy density lies mainly
in the gradient part of the dilaton field. From
Eq.  (\ref{eqn14}) it
follows that energy flux density is independent
of $\xi$:
\begin{equation}
\label{eqn15}
T_{0}^{0}=H_{0}^{2}(1+u^2); \quad T_{0}^{1}=2 H_{0}^{2} u.
\end{equation}
This can be interpreted as follows: the conversion of the
 energy of the  dilaton field from the gradient to the potential part
implies  a  conversion
of the gradient energy of the dilaton field to the energy of
electromagnetic field. Certainly,
this conversion is completely reversible.

\section{Dissipation}
\label{@DISS}
The above electro-dilaton soliton
appears at the front
 of the detonation wave. If we use the plasma behind the front
as a  frame of reference,   then this soliton will be move
with a velocity $u=c_s=1/\sqrt{3}$ in the positive $x$-direction.
Such a movement
generates  heat in the plasma and creates of pairs of particles
and antiparticles  in  high concentration.
This reduces the ${\bf E}$ and
${\bf H}$ fields. In the simplest description, the decay
wave can be taken into  account
in a travelling-wave approximation
by replacing  Eqs.~(\ref{eqn11})    by
\begin{equation}
\label{eqn16}
(1-u^2)\frac{d}{d\xi}(e^{-2\alpha\Phi}H_z)=\frc e^{-2\alpha\Phi}H_z; \quad
(e^{-2\alpha\Phi}H_z)=H_0 \exp[\frc \xi /(1-u^2)],
\end{equation}
where $\frc$ is the friction constant
with the dimension of a reverse length.
 While moving though the plasma with $\xi<0$,
the electromagnetic wave decays.
The equation for the dilaton  field with $\zeta=1$ becomes
\begin{eqnarray}
\frac{d^2\Phi}{d\xi^2}=-\frac{\alpha}{2}H_{z}^{2}e^{-2\alpha\Phi}.
\label{@}\end{eqnarray}
Substitution here (\ref{eqn16}) and using
the redefinitions (\ref{eqn13}), we  obtain
\begin{eqnarray}
\label{eqn17}
H_z&=&H_0 h(\eta) e^{\beta \eta},~~~~ {\rm
 with}~\quad \beta=\frc /2\alpha H_0,  \\
\frac{d^2 \psi}{d \eta^2}&
=&-\frac{1}{2}h^2(\eta)e^{-\psi}=-\frac{1}{2}e^{\psi+2\beta\eta}.
\end{eqnarray}
After a change of the variable $\psi+2\beta\eta\rightarrow\psi$, this equation
reduces to the  Liouville form (\ref{eqn9}).
Then subject to dissipation we have the following solution:
\begin{equation}
\label{eqn18}
\psi_d=-2[\beta\eta+\ln\cosh(\eta/2)]; \quad
H_d(\eta)=H_0 e^{-\beta \eta}/\cosh^2(\eta/2).
\end{equation}
Here, the subscript $d$ denotes a solution with dissipation. Thus electro-dilaton
 solitons
with dissipation become asymmetric with a  steeper front part. The creation of
particles and heat
 of plasma happens at decreasing of the energy density of  dilaton field.
For $|\eta|\rightarrow\infty$ the energy of the system is
concentrated only in the gradient part  of  dilaton field
(see (\ref{ener}),(\ref{redef}))
$$T_{0}^{0}=(1+u^2)H_{0}^{2}(\psi_{d,\eta})^2.$$
In the limit $|\eta|\rightarrow\infty$,
the equations (\ref{eqn18}) becomes
$$T_{0}^{0}(\infty)=(1+u^2)H_{0}^{2}(1+2\beta)^2; \quad
T_{0}^{0}(-\infty)=(1+u^2)H_{0}^{2}(1-2\beta)^2.$$
It is clear from here that the energy density of  dilaton
 field before the wave is more than the energy density behind the wave
($\beta>0$).
The limiting value for $\beta$ in this example is $\beta\rightarrow 0.5$.
This implies that the whole energy of the dilaton field transforms to the
heat energy of plasma.

\section{Induced Current}

Consider now the dissipation of energy by the induced fields
at the wave front. For this purpose we substitute a current
on the right hand side of Eq. (\ref{eqn11})
$$
j_y=\sigma E_y=\sigma u H_z(\xi),
$$
where $\sigma$ is specific conductivity of the medium.
Then we find the following system of dimensionless equations:
\begin{equation}
\label{eqn19}
2\frac{d^2\psi}{d\eta^2}=-h^2 e^{-\psi}, \quad \frac{d}{d\eta}(h e^{-\psi})=-\beta h,
\end{equation}
with the parameter $\beta\sim\sigma u$. At $\beta=0$, this reduces to the
previous   electro-dilaton soliton (\ref{eqn14}).

\begin{figure}[tbhp]
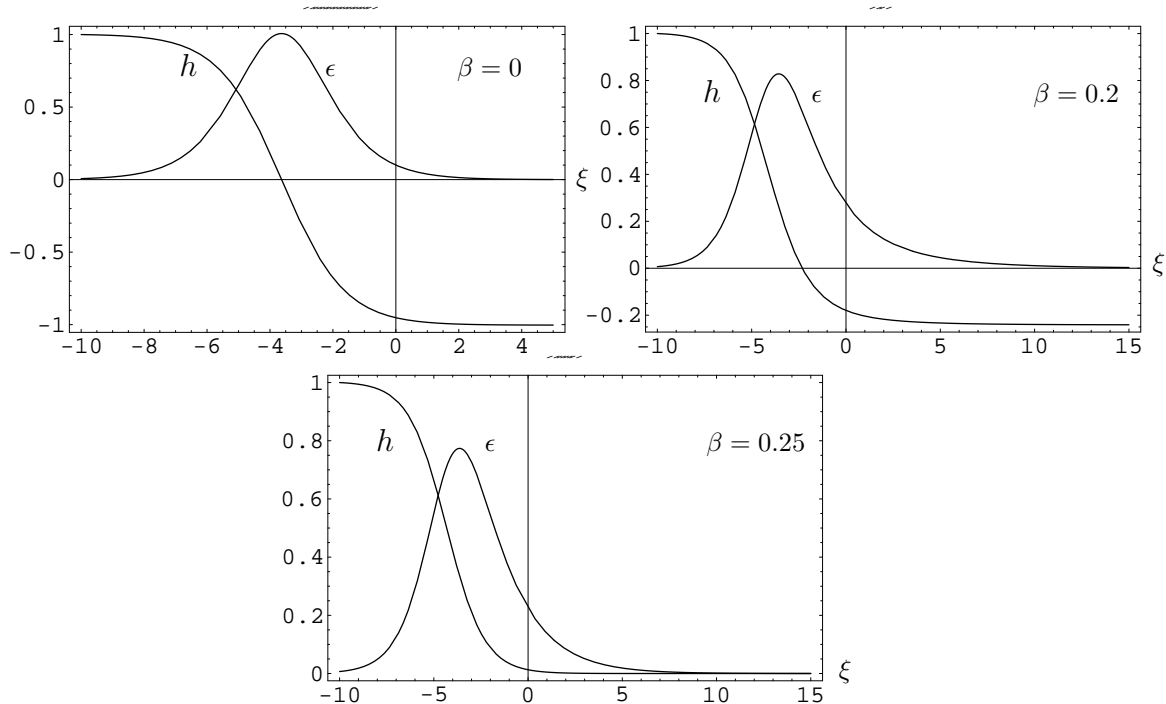

\input  b0.tps
~~~~~~~~\raisebox{-3.5mm}{\input  b02.tps}\\ [4mm]
~~~~~~~~~~~~~~~~~~~\phantom{xxxxxxx}~~~~~~~\raisebox{-6.5mm}{\input  b025.tps}
\caption[]{
Dependence of the energy density $\varepsilon$
and velocity $v$ of plasma on the self-similar variable $\xi$
behind the front of detonation wave
for various parameters $ \beta $.
}
\label{@Fig2}\end{figure}

To analyze Eqs. (\ref{eqn19}) it is convenient to introduce
the new variable
\begin{equation}
\label{eqn20}
z=h e^{-\psi},
\end{equation}
which is unity in the absence of dissipation.
The first integral  of the differential equations (\ref{eqn19})
leads to the  solution:
$$z^2=4\beta\left(\frac{\partial \psi}{\partial \eta}\right)
+C; \quad C=1-4\beta.$$
The integration constant is selected so that
in the limit $\eta\rightarrow -\infty$ the solution tends
to  the dissipation-free soliton with
$(d\psi/d\eta)_{-}\rightarrow 1$ for $z^2\rightarrow 1$.
In the opposite limit $\eta\rightarrow \infty$,
 the variable $z^2$ tends to zero  and
\begin{equation}
\label{eqn21}
\left(\frac{d\psi}{d\eta}\right)_{+}\rightarrow \frac{4\beta -1}{4\beta}.
\end{equation}
Thus for $z^2\rightarrow 0$, the solution represents a
kink   moving in  the positive $x$-direction.
By virtue of $\beta>0$, the asymptotic value of the dilaton
energy density  before  the front   is less than  behind it.
 Physically this  means that the pressure of the dilaton field
creates a plasma at the wave front pushing it
ahead. It follows from Eq. (\ref{eqn21})  that the
limiting value of $\beta$ is now $\beta_c=1/4$
implying that the electro-dilaton energy
goes completely over into an energy flux of
moving plasma.  The solutions are plotted in Fig.~\ref{@Fig2}.

\section{Conclusion}

>From our discussion   it appears perfectly plausible that
dark matter consists  of electro-dilaton stars and is not dark at all,
but has been showing its presence quite dramatically
all along via the powerful gamma-ray bursts
observed since.

\section{Acknowledgement}
VF and VG than the DAAD in Bonn/Germany for financial support.
We also thank Dr. A. Pelster for discussions.

\end{document}